\def\year{2020}\relax
\documentclass[letterpaper]{article} 
\usepackage{aaai20}  
\usepackage{times}  
\usepackage{helvet} 
\usepackage{courier}  
\usepackage[hyphens]{url}  
\usepackage{graphicx} 
\usepackage{graphicx}  
\usepackage{amsmath}
\usepackage{booktabs}
\usepackage{tikz}
\usetikzlibrary{positioning,chains,fit,shapes,calc}

\newcommand{\var}{\mathit{var}}

\newcommand*\circled[1]{\tikz[baseline=(char.base)]{
            \node[shape=circle,draw,inner sep=2pt] (char) {#1};}}

\begin{document}

\pdfinfo{
/Title (Constructing Minimal Perfect Hash Functions Using SAT Technology)
/Author (Sean A. Weaver and Marijn J.H. Heule)}

\title{Constructing Minimal Perfect Hash Functions Using SAT Technology}


\author{
Sean A. Weaver\\
Laboratory for Advanced Cybersecurity Research\\
           U.S. National Security Agency\\
           {saweave@evoforge.org}
\And
Marijn J.H. Heule\\
Computer Science Department\\
Carnegie Mellon University\\
{marijn@cmu.edu}\\
Amazon Scholar
}
%

\maketitle

\begin{abstract}

Minimal perfect hash functions (MPHFs) are used to provide efficient
access to values of large dictionaries (sets of key-value
pairs). Discovering new algorithms for building MPHFs is an area of
active research, especially from the perspective of storage
efficiency. The information-theoretic limit for MPHFs is $\frac{1}{\ln
  2} \approx 1.44$ bits per key.  
The current best practical algorithms range between 2 and 4 bits per key. 
In this article, we propose two SAT-based constructions of MPHFs.
Our first construction yields MPHFs near the information-theoretic limit.
For this construction, current state-of-the-art SAT solvers can handle instances where the dictionaries contain up to 40 elements,
thereby outperforming the existing (brute-force) methods.
Our second construction uses XORSAT filters to realize a practical approach with
long-term storage of approximately $1.83$ bits per key. 
\end{abstract}



\section{Introduction}\label{sec:Intro}

A \emph{minimal perfect hash function} (MPHF) for a set $Y$ with $n$ distinct elements
is a collision-free mapping from the elements of $Y$ to the set $[n] = \{1,\dots,n\}$. MPHFs enable
efficient access to data stored in large databases by providing a
unique index for each key in a set of key-value
pairs. This allows the value of each key-value pair to be stored at its associated index in an $n$-entry table. Since industrial databases
are often of significant size, if an MPHF is going to be used, one would
want it to use as little extra space as possible. 

The information-theoretic limit for MPHFs is \mbox{$\frac{1}{\ln 2} \approx
1.44$} bits per key~\cite{Mehlhorn82,Fredman84}, yet no practical
constructions (those not relying on brute-force) have been found that meet
this limit. In fact, the current best algorithms range between 2 and 4 bits per
key~\cite{Belazzougui09,Botelho13,Genuzio16,Limasset17}. 
Since applications may also consider costs such as build time and
query time, MPHF construction and querying should not be so costly as to
outweigh the benefit of a compact representation. 

We introduce two new constructions for static MPHFs: i.e.,
MPHFs where the set $Y$ is immutable and known in advance. Both constructions are based
on satisfiability (SAT) techniques and utilize a universal family of hash functions. 

Our first construction encodes the MPHF construction as a Boolean formula
and computes a solution for it using a SAT solver. 
It aims at the construction of MPHFs near
the information-theoretic limit (roughly $1.44$ bits per key for large sets, say consisting of over a few thousand keys) and
can be queried in $\log_2 n$ steps. The information-theoretic limit $\alpha_n$ for smaller values of $n$ are shown in
Table~\ref{tab:limit}.

We present a compact $\mathcal{O}(n^2 \log_2n)$ SAT encoding and a more effective
$\mathcal{O}(n^3)$ encoding. 
On these encodings, current SAT solvers (both complete and incomplete approaches) exhibit exponential runtime
on the number of elements in the dictionary, with runtimes growing
prohibitively large around $40$ elements, making our construction
currently impractical for large sets. However, to the best of our knowledge,
no other existing construction (including brute-force) can construct MPHFs
near the information-theoretic limit for sets with more than $20$ elements
in reasonable time.

Since we are only interested in
satisfiable formulas (existence of MPHFs), SAT approaches based on
local-search technology seem a natural fit. Currently complete solving 
methods work best on our encodings. However, our construction creates formulas with a
strong random flavor and enormous progress has been made 
on local-search techniques for uniform random $k$-SAT instances~\cite{KautzS96,BalintS12,Gableske14}.

Our second construction uses approximately $1.83$ bits per key and can build
MPHFs in $\mathcal{O}(n^3)$ steps (though construction is trivially
parallelizable). These MPHFs can be queried in $max(3, \ln n + \ln(\ln
n))$ steps in the worst case. Formally, the construction involves storing
a minimum-weight perfect matching of a weighted bipartite graph in a
space-efficient retrieval structure called an XORSAT
filter~\cite{Weaver18}. Our implemented prototype demonstrates
that the approach can produce MPHFs for large datasets ($2^{30}$ keys)
with fast query speed (a million queries per second) for low costs. 


\begin{table}[t]
\caption{The information-theoretic limit ($\alpha_n$) of bits per key for a various number of keys ($n$).}
\label{tab:limit}
\begin{center}
\begin{tabular}{@{}c|c@{~~\,}c@{~~\,}c@{~~\,}c@{~~\,}c@{~~\,}c@{~~\,}c}
$n$  & $10$ & $20$ & $30$ & $40$ & $10^2$ & $10^3$ & $10^4$ \\ \midrule
$\alpha_n$ & $1.143$ & $1.268$ & $1.317$ & $1.343$ & $1.396$ & $1.436$ & $1.442$
\end{tabular}
\end{center}
\end{table}

\section{Preliminaries}

Below we present the most important background concepts related to the contributions of this paper.

\paragraph{Minimal Perfect Hash Functions.}

A \emph{minimal perfect hash function} (MPHF) for a set $Y$ with $n$ distinct elements 
is a bijection that maps the elements from $Y$ to the set $[n] = \{1,...,n\}$.
Three important tradeoffs play a role when constructing MPHFs:
{\em storage space}, {\em query speed}, and {\em
building cost}. Most research focuses on lowering the storage space,
while having acceptable query speed and building cost. 

The probability that a universal function $H$ with range $[n]$ is a minimal perfect hash function
for $Y$ is $\frac{n!}{n^n}$. For each element $y_i$ with $i \in [n]$, the probability
that $H(y_i)$ and $H(y_j)$ with $(j<i)$ don't collide is $\frac{n+1-i}{n}$. 
The information theoretic limit $\alpha_n$ is ${\log_2 (\frac{n^n}{n!})}/n$ bits per element.
This limit is roughly $1.44$ bits per element for large $n$, but 
smaller for small $n$. Table~\ref{tab:limit} shows the limit for various values of $n$.

The only existing MPHF constructions that realize this bound are
based on brute-force: they evaluate many hash functions on $Y$ and terminate
as soon as a minimal perfect hash function is found. As the representation of such a hash function
requires on average $\alpha_n$ bits per element, it follows that, for example, brute-force over a set of $20$ elements
requires on average evaluating over 43 million ($2^{25.36}$) hash functions.

\paragraph{Propositional Logic and Satisfiability.}
We consider propositional formulas in \emph{conjunctive normal form}, 
which are defined as follows. A \emph{literal} is either a variable $x$ (a \emph{positive literal}) 
or the negation $\overline x$ of a variable~$x$ (a \emph{negative literal}). 
For a literal $l$, we denote the variable of $l$ by $\var(l)$. 
The \emph{complementary literal} $\overline l$ of a literal $l$ is defined as 
$\overline l = \overline x$ if $l = x$ and $\overline l = x$ if $l = \overline x$.
A \emph{clause}  is a finite disjunction of the form $(l_1 \lor \dots \lor l_k)$, where $l_1, \dots, l_k$ are literals.
A \emph{formula} is a finite conjunction of the form $C_1 \land \dots \land C_m$, where $C_1, \dots, C_m$ are clauses.
An \emph{XOR constraint} is an expression of the form $l_1 \oplus \dots \oplus l_k \equiv b$, where $l_1, \dots, l_k$ are literals and $b \in \{0,1\}$. 

An \emph{assignment} is a function from a set of variables to the truth values ${\tt 1}$~(\emph{true}) and 
${\tt 0}$~(\emph{false}).
A literal $l$ is \emph{satisfied} by an assignment $\alpha$ if 
$l$ is positive and \mbox{$\alpha(\var(l)) = 1$} or if it is negative and $\alpha(\var(l)) = 0$.
A literal is \emph{falsified} by an assignment if its complement is satisfied by the assignment.
For a literal $l$, we sometimes slightly abuse notation and write $\alpha(l) = 1$ if $\alpha$ satisfies $l$ and $\alpha(l) = 0$ if $\alpha$ falsifies $l$.
A clause is satisfied by an assignment $\alpha$ if it contains a literal that is satisfied by~$\alpha$.
Finally, a formula is satisfied by an assignment $\alpha$ if all its clauses are satisfied by $\alpha$.
A formula is \emph{satisfiable} if there exists an assignment that satisfies it and \emph{unsatisfiable} otherwise.
Moreover, an XOR constraint $l_1 \oplus \dots \oplus l_k \equiv b$ is satisfied by an assignment $\alpha$ if $\alpha(l_1) + \dots + \alpha(l_k) \equiv b\ (\text{mod}\ 2)$.

\paragraph{XORSAT filters.}

An XORSAT filter is a space-efficient probabilistic data structure
used for testing whether an element is in a set. In the context of this
paper, XORSAT filters are treated as dictionaries of one-bit items. An
XORSAT filter consists of $n$ bits and $k$ hash functions with range
$[n]$. To retrieve the stored bit associated with an element, $k$ hash
functions are evaluated on the element. This results in $k$ entries in
the array. If the number of entries with value {\tt 1} is even, then the
stored bit is {\tt 0}, otherwise the stored bit is {\tt 1}. XORSAT
filters are constructed as follows: First, an XOR constraint 
of length $k$ is generated for each of the $n$ elements in the set, with the right-hand side of
each constraint being equal to the corresponding bit to be stored. 
After this, the XORSAT filter is constructed by computing a
solution for the conjunction of these XOR constraints.

\section{A Space-Optimal MPHF Construction}

We first explore the ability of SAT solving techniques to construct
space-optimal ($\alpha_n$ bits per key) MPHFs.
We assume given a set $Y = \{y_1,\dots,y_n\}$ (the elements to be hashed) and an integer $m$ (the number of bits used for storing the MPHF). 
Clearly, the elements of $Y$ can be represented by bit-vectors of length $k = \lceil \log_2(n)\rceil$. 
For our constructions, we use a universal family of $k$ hash functions $H_1, \dots, H_k$ with domain $Y$ and range $\{-m,\dots,m\} \setminus \{0\}$. 
For an element $y_j$, we denote with $L_i(y_j)$ the literal $x_{H_i(y_j)}$ if $H_i(y_j) > 0$ and the literal $\overline x_{-H_i(y_j)}$ otherwise.

To encode the problem of finding MPHFs into propositional logic,
we first introduce $m$ distinct Boolean variables $x_1,\dots, x_m$. 
We then aim at defining a propositional formula such that the index (the hash) 
for each element $y_i$ can be derived from a satisfying assignment $\alpha$ of the formula (and
in particular from the truth values assigned to $x_1, \dots, x_m$) as follows:
The index for $y_i$ is going to be the bit-vector $\alpha({L_1(y_j)}) \dots \alpha({L_k(y_j)})$. 
For example, let $m = 3$, $j = 1$, and assume that $H_1(y_1) = 3$, $H_2(y_1) = -1$, and $H_3(y_1) = 2$.  
Then, the index of $y_1$ is determined by the truth values assigned to the literals ${L_1(y_1)} = x_3$, 
${L_2(y_1)} = \overline x_1$, and ${L_3}(y_1) = x_2$. Thus, if $\alpha(x_1) = 1$, $\alpha(x_2) = 0$ and $\alpha(x_3) = 1$, 
then $y_1$ will be mapped to the index $\alpha(x_3)\alpha(\overline x_1)\alpha(x_2) = \langle{\tt 100}\rangle = 5$ 
(note that in our setting the bit-vector $\langle{\tt 100}\rangle$ represents the number $5$ instead of $4$ since indices start at 1 and not at 0). 

To ensure that the mapping of each element depends on $k$ distinct variables, we enforce that each hash function returns a different variable by applying a hash function multiple times in case of a clash.  

To obtain an MPHF, we want a propositional formula that is satisfied by an assignment $\alpha$ if and only if 
%
for every pair of distinct elements
$y_i$ and $y_j$, it holds that $\alpha({L_1(y_i)}) \dots \alpha({L_k(y_i)}) \neq \alpha({L_1(y_j)}) \dots \alpha({L_k(y_j)})$; 
in other words, it holds that $y_i$ and $y_j$ are mapped to different indices.
Once we obtain such an assignment, we transform it into a bit-vector (the storage) by concatenating the truth-value assignments to all the literals. 
From this bit-vector we can then compute the index of an element bit-by-bit via simple look-ups
using the hash functions $H_1,\dots,H_k$. 

\paragraph{Example 1.} Let $Y$ be the set $\{y_1, y_2, y_3, y_4\}$,
resulting in $n = 4$ and $k=\lceil \log_2(n)\rceil = 2$. For this
example, we choose the number of Boolean variables to be $m=6$. 
Suppose the evaluation of the $k=2$ hash functions $H_1$ and $H_2$ 
(with range $\{-m,\dots,m\} \setminus \{0\}$) on all elements of $Y$ produces the following table:

\begin{center}
{\large
\begin{tabular}{c|cc}
& $L_1$ & $L_2$\\
\midrule
$y_1$ & $\overline x_1$ & $x_5$ \\
$y_2$ & $\overline x_2$ & $\overline x_6$ \\
$y_3$ & $x_3$ & $x_1$\\
$y_4$ & $\overline x_5$ & $x_4$\\
\end{tabular}}
\end{center}

The resulting bit-vectors are then obtained from the assignments to the variables in 
$\overline x_1x_5$ for $y_1$, $\overline x_2 \overline x_6$ for $y_2$, $x_3x_1$ for
$y_3$, and $\overline x_5x_4$ for $y_4$. 
For instance, the assignment that maps the truth value ${\tt 0}$ to the variables $x_2$ and $x_4$, and the truth value ${\tt 1}$ to the other variables 
makes all indices different: $y_1$ maps to $2$
($\langle{\tt 01}\rangle$), $y_2$ to $3$ ($\langle{\tt 10}\rangle$), $y_3$ to $4$ ($\langle{\tt 11}\rangle$), and
$y_4$ to $1$ ($\langle{\tt 00}\rangle$). The storage of the MPHF is $\langle{\tt 101011}\rangle$: 6
bits for 4 elements, thus 1.5 bits per element. A query to obtain the
value for an element requires $k$ look-ups into the stored bit-vector.

\begin{figure*}[t]
\begin{minipage}{.48\textwidth}
\includegraphics{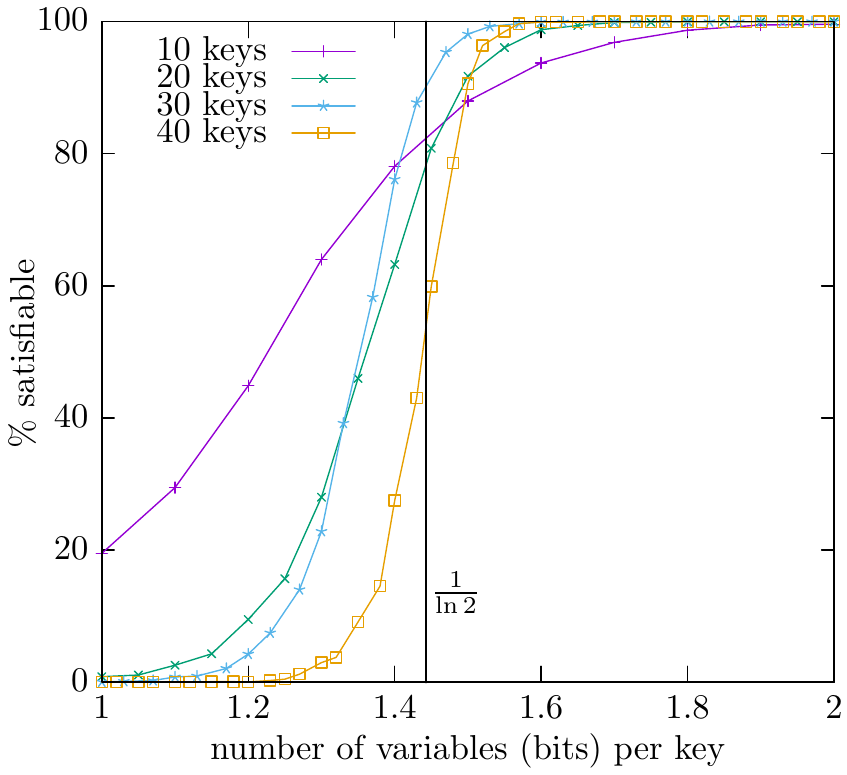}
\end{minipage}
~~
\begin{minipage}{.48\textwidth}
   \includegraphics{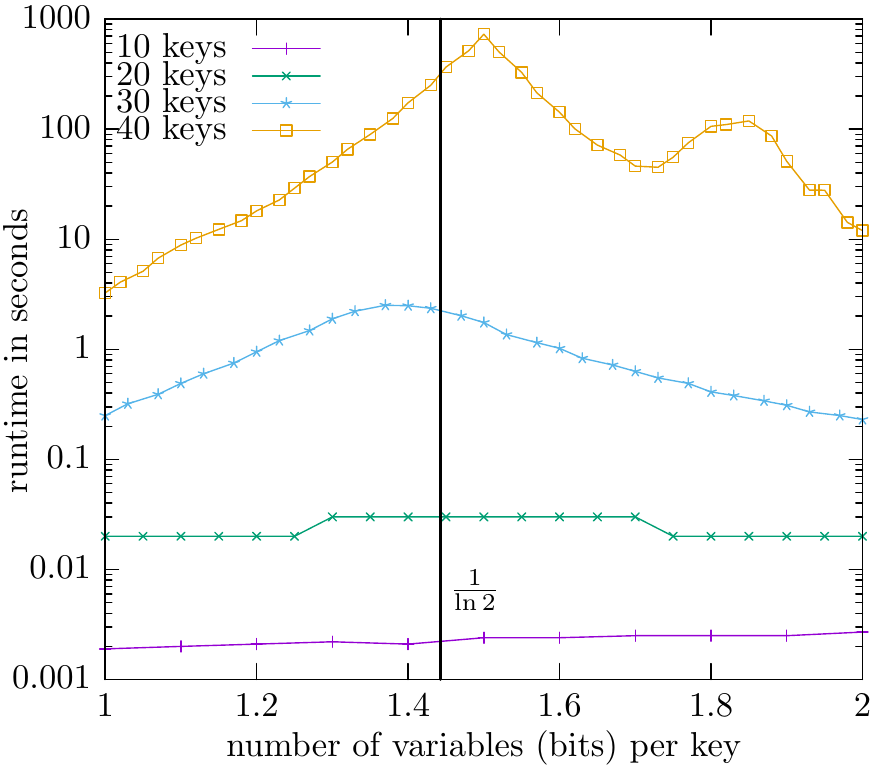}
\end{minipage}
    \caption{Left, the satisfiability threshold for MPHFs with $10$, $20$, $30$, and $40$ keys.
      The horizontal axis shows the number of bits per key. The
      vertical axis is the fraction of runs (out of $2400$) that were
      satisfiable. The bold vertical line represents the MPHF
      information-theoretic limit for large sets. Right, the average runtime to build MPHFs with $10$, $20$, $30$, and $40$ keys.
      The vertical axis shows the runtime in seconds using a logscale.}\label{fig:threshold}
\end{figure*}

\subsection{Encodings}

The encoding of a problem into SAT can have a big impact on the
performance of SAT solvers. We observed this for MPHF as well. 
An often used heuristic for an effective encoding is to minimize
the sum of the number of variables and the number of clauses.  
This heuristic is, for example, used in the most-commonly applied SAT-preprocessing technique---\emph{bounded variable elimination}~\cite{Een05}. We thus 
focused our initial efforts on designing a compact encoding.

If $n$, the number of keys, is a power of 2, then the encoding only requires
the following \emph{all-different} constraint, which we need to
encode in conjunctive normal form:
\begin{equation}
\label{eq:alldiff}
\bigwedge_{1\leq i<j\leq n} 
{L_1(y_i)} \dots {L_k(y_i)} \neq {L_1(y_j)} \dots {L_k(y_j)}.
\end{equation}
Additionally, if the number of keys is not a power of 2, then we need
a constraint stating that none of the elements should be mapped to values larger than $n$.

The first and most compact encoding of the all-different constraint
used in our experiments introduces auxiliary variables $e_{a,b}$
with $a, b \in \{x_1, \overline x_1, \dots, x_m, \overline x_m\}$ that denote that literals $a$ and
$b$ are equal. So $e_{a,a} = 1$,  $e_{a,\overline a} = 0$, and $e_{a,b} = e_{b,a}$. 
These variables are used to construct the clauses
\begin{equation}
\bigwedge_{1\leq i<j\leq n} (\overline e_{L_1(y_i),L_1(y_j)} \lor \dots \lor \overline e_{L_k(y_i), L_k(y_j)}).
\end{equation}
Intuitively, these clauses state that the keys for distinct elements $y_i$ and $y_j$ need to differ in at least one bit.

Notice that we only need such $e_{a,b}$ variables if for some pairs of elements $y$ and $y'$ and some
$i$ holds that $L_i(y) = a$ and $L_i(y') = b$.
The intuitive meaning of the $e_{a,b}$ variables can simply be encoded using an XOR
constraint:
\begin{equation}
e_{a,b} \oplus a \oplus b \equiv 1.
\end{equation}
Although encoding ternary XOR constraints normally requires four
clauses, we used only the two clauses with the positive literal
$e_{a,b}$ as the two clauses with the negative literal $\overline e_{a,b}$
are \emph{blocked}~\cite{Kullmann99} and thus redundant. 
The resulting encoding has $\mathcal{O}(n^2 \log_2(n))$ ternary clauses and 
$\mathcal{O}(n^2)$ clauses of length $k$.

We also explored various other encodings since the results based on the
above encoding were disappointing. The most effective encoding (for
both complete and incomplete solvers) that we explored uses no
auxiliary variables---in case the number of keys is a power of 2, we
encode the constraint~(\ref{eq:alldiff}) as follows:

\begin{equation}
\label{eq:l}
\!\bigwedge_{~1\leq i,j,l \leq n, i \neq j\!\!\!\!} \!\!\!\!\!\!\!\!\!\! 
{L_1(y_i)} \dots {L_k(y_i)} = l \rightarrow {L_1(y_j)} \dots {L_k(y_j)} \neq l.\!\!
\end{equation}

In the above, ${L_1(y_i)} \dots {L_k(y_i)} = l$  
denotes that the bit-vector interpretation of  ${L_1(y_i)} \dots {L_k(y_i)}$, with ${L_1(y_i)}$
being the most significant bit, equals the $l$th position in range $[n]$. E.g.,
$\langle{\tt 000}\rangle = 1$ and $\langle{\tt 001}\rangle = 2$. In case the number of keys is not a power of 2,
we drop both literals that represent the most significant bits, i.e., $L_1(y_i)$ and $L_1(y_j)$, 
for all $l$ with $l + 2^{k-1} > n$ and $l - 2^{k-1} < 1$. 

When representing this encoding in conjunctive normal form, 
the number of clauses is cubic in $n$ and therefore
much larger compared to the first encoding, which uses auxiliary variables.
Also, the clauses are of length $2k$. 
It is therefore unlikely that one can use this encoding to 
construct MPHFs with hundreds of keys because solving almost any formula
with over a million clauses is challenging for current state-of-the-art solvers.
The encoding also produces many tautological clauses and duplicate literals, which
are both discarded by our generator. 

The cubic encoding turned out to be more effective compared to the 
compact encoding. For example, formulas with 32 keys and 46 variables
can be solved with the cubic encoding with an average runtime of 4 seconds
whereas the average solving time for formulas based on the compact encoding was
8.5 seconds. One explanation for the difference is that the cubic encoding
enables more unit propagations.
Consider again Example 1 and in particular the constraint stating that the bit-vectors of $y_1$
and $y_3$ must be different. 
In terms of variables it means that $\overline x_1x_5$ cannot be equal to $x_3x_1$. 
Under the assignment that makes $x_3$ true and $x_5$ false, 
we can deduce that $x_1$ must be true. Unit propagation on the cubic 
encoding will infer this. However, in the compact encoding we end up 
with the binary clauses $(\overline e_{x_1,x_3} \lor \overline e_{\overline x_1,x_5})$,
$(e_{x_1,x_3} \lor x_1)$, and $(e_{\overline x_1,x_5} \lor x_1)$.



Other encodings of the all-different constraint of bit-vectors have been
studied in the literature~\cite{Biere08AllD,Surynek12}, 
but the authors considered the case where the number of bit-vectors is significantly smaller than the range of the bit-vectors. For our application, however, the number of bit-vectors
and their range is equal or similar.

We observed that SAT encodings of MPHF problems with
$m \geq \lceil n / \ln 2 \rceil$ have a high probability of
being satisfiable (in the range of $n$ up to 40 used in the
evaluation) and that a phase transition pattern can be observed near
the information-theoretic limit of constructing MPHFs. We experimented with a
range of different SAT encodings and SAT solvers, but the current
state-of-the-art tools are not powerful enough to practically construct MPHFs with
more than 40 elements near the theoretic limit.

       


\subsection{Experimental Evaluation}

We evaluated the effectiveness of our SAT-based approach to constructing
MPHFs near the information-theoretic limit. The encoding and decoding tools are 
available at {\tt https://github.com/weaversa/MPHF}. We used cube-and-conquer~\cite{CnC} to solve the resulting formulas
as this approach demonstrated strong performance on the hardest MPHF instances. 
Cube-and-conquer splits a given problem into multiple subproblems, which are solved by a ``conquer'' solver. We used
{\tt glucose} 3.0~\cite{Audemard:2009} as conquer solver. All mentioned runtimes are the sum of splitting the problem and
the solving time by the conquer solver on a single core of a Xeon CPU E5-2650.

Figure~\ref{fig:threshold} (left) shows the results for 
$10$, $20$, $30$, and $40$ random keys with the number of bits per key between 1 and 2. 
Each point shows the fraction of satisfiable formulas averaged over 2400 instances
for each number of keys ($n$) and number of variables ($m$). The bold vertical 
line denotes the information-theoretic limit. 

The observed phase transition from UNSAT to SAT becomes 
more and more sharper when increasing the number of keys. 
Moreover, the curve based on 40 keys appears to be almost 
symmetric (rotation by 180 degrees) in the point at 50\% at the 
information-theoretic limit for large $n$. 
Recall that the information-theoretic limit $\alpha_{40}=1.343$. Hence the 
phase transition happens slightly after the limit. We analyzed the results and
noticed various instances that did not use all $m$ variables. If one or more variables
are missing, then the formula is more likely to be unsatisfiable. The observed phase transition
for $n=40$ is much closer to $\alpha_{40}$ when only
considering the instances that use all $m$ variables.

Our results show that the SAT-based 
approach is able to construct close to space-optimal MPHFs for 
the limited number of keys considered in the experimental evaluation. 
They also suggest that the curve would further sharpen when increasing
the number of keys. To the best of our knowledge, this is the first 
approach that can construct close to space-optimal MPHFs for 40 keys in
reasonable time. 

Unfortunately, experiments with off-the-shelf SAT
solvers on the encodings exhibit exponential runtime on the number of
keys as shown in Figure~\ref{fig:threshold} (right). The runtime bump near 1.8 for $40$ keys is
due to the lack of subproblems that are generated. Generating more subproblems for these 
instances would reduce the runtime (combined splitting and solving time). 
The formulas for 40 keys at the observed phase transition require on average about 800 seconds, although for some instances solving takes a few hours.
In contrast, the hardest instance for 30 keys can be solved in a several seconds. 
Hence, larger runs that may sharpen the logistic curve could
not be completed. We experimented with both complete and incomplete state-of-the-art solvers, including {\tt probsat}~\cite{BalintS12}. Whether there exists
an encoding that is more amenable to SAT solvers is left for future work. 

\section{A $1.83n$ MPHF Construction}\label{sec:Building}

The second MPHF construction, which allows building MPHFs with
approximately $1.83n$ bits per key, also uses a universal family of $k$ hash
functions and returns a bit-vector of $m$ bits. However, obtaining the
index for a key is quite different. Instead of computing the index bit
by bit, the second MPHF approach computes which hash function should
be used to obtain the index.

To build an MPHF for a set $Y$ with $n$ elements,
given a set of universal hash functions ${H_1, ..., H_k}$ each with range $[n]$, 
we first create a weighted bipartite graph $G = (Y, [n], E)$, 
where for all $y \in Y$ and for all $i \in
[k], (y, H_i(y)) \in E$, it holds that $\mathit{weight}(y, H_i(y)) = i$.

Here, the weighted bipartite graph is created in such a way that the
nodes on the left-hand-side (i.e., the left partition) 
represent elements of $Y$ and the right-hand-side
nodes represent elements from $[n]$. The hash functions
determine how every left-hand-side node is connected to $k$
right-hand-side nodes. That is, for each $y \in Y$, compute ${H_1(y),
\dots, H_k(y)}$ and add edges from $y$ to the resulting nodes
produced by the hash functions. The weight of each edge is assigned
the index of the hash function used, that is, $\mathit{weight}(y,
H_i(y)) = i$.

The parameter $k$ should be chosen in a way that guarantees that $G$
possesses a perfect matching. A result by Frieze and
Melsted~\cite{Frieze12} shows that, with high probability, $G$ will
have a perfect matching when the degree of each left-hand-side node is
at least three and the degree of each right-hand-side node is at least
two. Newman and Shepp's result~\cite{Newman60} on the {\em Double
Dixie Cup Problem} shows that the right-hand-side nodes will, with
high probability, have degree two when the number of edges reaches $n
(\ln n + \ln(\ln n))$. Hence, since every $H_i$ is used to hash $n$
keys, the optimal value for parameter $k$ is $max(3, \ln n + \ln(\ln
n))$.

Let $M$ be a minimum-weight perfect matching of $G$. $M$ can be found
in $\mathcal{O}(n^3)$ steps using the Hungarian
Algorithm~\cite{Kuhn55}. $M$ is also the bijective mapping of the MPHF
being built. Since $G$ fits Parviainen's proposed ``Case
II''~\cite{Parviainen04}, $M$ has weight approximately $1.83n$,
asymptotically, which was determined experimentally.

The mapping $M$ is stored in a space-efficient retrieval structure
such that, for each $y$, if $H_i(y) \in M$ then store $((i, y), 1)$
(definite presence) and for each $j < i$ store $((j, y), 0)$ (definite
absence).\footnote{This is reminiscent of BBHash's positional MPHF
scheme~\cite{Limasset17}} Such a retrieval structure can be created in
$\mathcal{O}((1.83n)^3)$ steps, takes up one bit of space per element
stored, and can be accessed in a small constant number of
steps~\cite{Dietzfelbinger19}. It's worth noting that both the
Hungarian Algorithm and the retrieval structure construction process
can benefit from sharding the input into small sets (which can all be
processed in parallel), meaning that the MPHF can be created more
efficiently than two $\mathcal{O}(n^3)$ steps, though at a
determinable, yet practically small loss to space efficiency. The
long-term storage of this MPHF construction is equivalent to the size
of the retrieval structure which will be approximately $1.83n$.

To find the mapping from $y$ to its corresponding index, query the
retrieval structure with $(i, y)$, starting with $i = 1$ and increment
until $1$ is given. Then, the index for $y$ is $H_i(y)$.

\paragraph{Example 2.} We provide a detailed example of building an MPHF by
storing a minimum weight perfect matching of a bipartite graph in the
solution of a $k$-XORSAT instance.\footnote{Many practical algorithms
  exist that support linear-time construction and constant-time
  look-ups on static
  dictionaries~\cite{Seiden94,Dietzfelbinger08,Aumuller09,Porat09,Botelho13,Genuzio16,Dietzfelbinger19}. Without
  loss of generality, the XORSAT filter~\cite{Weaver18} is chosen
  here.} Let $Y$ be the set $\{y_1, y_2, y_3, y_4,
y_5\}$, resulting in $n = 5$ and $k = max(3, \ln 5 + \ln(\ln 5)) =
3$. First, the $k$ hash functions are evaluated on all elements of
$Y$, producing the following table.

\begin{center}
{\large
\begin{tabular}{c|ccc}
& $H_1$ & $H_2$ & $H_3$\\
\midrule
$y_1$ & 1 & 5 & 2\\
$y_2$ & 2 & 4 & 5\\
$y_3$ & 1 & 3 & 4\\
$y_4$ & 1 & 3 & 1\\
$y_5$ & 5 & 3 & 3\\
\end{tabular}}
\end{center}

Next, a bipartite graph is built from the table by adding for each element in $Y$ an edge from the element to the results of the hash functions. Thus, for instance, $y_1$ is connected to $1$, $5$, and $2$. 

\begin{center}
{\large
\begin{tikzpicture}[thick,
  every node/.style={draw,circle,inner sep=1pt},
  fsnode/.style={fill=black},
  ssnode/.style={fill=black},
  every fit/.style={ellipse,draw,inner sep=-2pt,text width=2cm},
  -,shorten >= 3pt,shorten <= 3pt
  ]
  
\begin{scope}[start chain=going below,node distance=7mm]
\foreach \i in {1,2,...,5}
  \node[fsnode,on chain] (f\i) [label=left: $y_\i$] {};
\end{scope}

\begin{scope}[xshift=4cm,start chain=going below,node distance=7mm]
\foreach \i in {1,2,...,5}
  \node[ssnode,on chain] (s\i) [label=right: \i] {};
\end{scope}

\draw[very thick] (f1) -- (s1);
\draw (f1) -- (s5);
\draw[thin] (f1) -- (s2);
\draw[very thick] (f2) -- (s2);
\draw (f2) -- (s4);
\draw[thin] (f2) -- (s5);
\draw[very thick] (f3) -- (s1);
\draw (f3) -- (s3);
\draw[thin] (f3) -- (s4);
\draw[very thick] (f4) -- (s1);
\draw (f4) -- (s3);
\draw[very thick] (f5) -- (s5);
\draw (f5) -- (s3);
 
\end{tikzpicture}}
\end{center}

The weight of each edge is equal to the index of the hash function
used. For example, edge $(y_1, 1)$ has weight $1$ and edge $(y_2, 5)$
has weight 3. The Hungarian Algorithm is now used to find a minimal
weight perfect matching of the bipartite graph. One such matching
(which has weight $8$) is given below.

\begin{center}
{\large \centering
\begin{tabular}{c|ccc}
& $H_1$ & $H_2$ & $H_3$\\
\midrule
$y_1$ & \circled{1} & 5 & 2\\
$y_2$ & \circled{2} & 4 & 5\\
$y_3$ & 1 & 3 & \circled{4}\\
$y_4$ & 1 & \circled{3} & 1\\
$y_5$ & \circled{5} & 3 & 3\\
\end{tabular}}
\end{center}

This corresponds to the following matching:

\begin{center}
{\large
\begin{tikzpicture}[thick,
  every node/.style={draw,circle,inner sep=1pt},
  fsnode/.style={fill=black},
  ssnode/.style={fill=black},
  every fit/.style={ellipse,draw,inner sep=-2pt,text width=2cm},
  -,shorten >= 3pt,shorten <= 3pt
  ]
  
\begin{scope}[start chain=going below,node distance=7mm]
\foreach \i in {1,2,...,5}
  \node[fsnode,on chain] (f\i) [label=left: $y_\i$] {};
\end{scope}

\begin{scope}[xshift=4cm,start chain=going below,node distance=7mm]
\foreach \i in {1,2,...,5}
  \node[ssnode,on chain] (s\i) [label=right: \i] {};
\end{scope}

\draw[very thick] (f1) -- (s1);
\draw[very thick] (f2) -- (s2);
\draw[thin] (f3) -- (s4);
\draw (f4) -- (s3);
\draw[very thick] (f5) -- (s5);
  
\end{tikzpicture}}
  
\end{center}

The final step is to store the matching in an XORSAT filter. For this
example, we store the following $8$ elements:

\begin{center}
\begin{tabular}{ccc}
  $((y_1, 1), 1),$\\
  $((y_2, 1), 1),$\\
  $((y_3, 1), 0),$ & $((y_3, 2), 0),$ & $((y_3, 3), 1),$\\
  $((y_4, 1), 0),$ & $((y_4, 2), 1),$\\
  $((y_5, 1), 1)$.
\end{tabular}
\end{center}
Elements are stored in the filter in the way proposed by Weaver et
al.~\cite{Weaver18}, where the first part of the tuple is the element
being filtered on and the second part is one bit of metadata to store
(treating the filter like a dictionary of one-bit items).

Let $t$ denote the number of tuples, in this case 8. 
We construct the XORSAT filter using two hash functions with range $[t]$.
We limit the number of hash functions for readability, but use 5 hash
functions in practice to generate an XORSAT filter with high
efficiency.

\begin{center}
{\large
\begin{tabular}{c|cc|c}
tuple & $H_1$ & $H_2$ & XOR constraint\\
\midrule
$(y_1, 1)$  & 3 & 6 & $x_3 \oplus x_6 \equiv 1$\\
$(y_2, 1)$  & 5 & 1 & $x_5 \oplus x_1 \equiv 1$\\
$(y_3, 1)$  & 4 & 8 & $x_4 \oplus x_8 \equiv 0$\\
$(y_3, 2)$  & 2 & 3 & $x_2 \oplus x_3 \equiv 0$\\
$(y_3, 3)$  & 5 & 4 & $x_5 \oplus x_4 \equiv 1$\\
$(y_4, 1)$  & 8 & 7 & $x_8 \oplus x_7 \equiv 0$\\
$(y_4, 2)$  & 2 & 7 & $x_2 \oplus x_7 \equiv 1$\\
$(y_5, 1)$  & 4 & 3 & $x_4 \oplus x_3 \equiv 1$\\
\end{tabular}}
\end{center}

The conjunction of the XOR constraints is satisfiable, for example
by assigning the truth value ${\tt 1}$ to the variables $x_2$, $x_3$, and $x_5$, 
and the truth value ${\tt 0}$ to the other variables. We thus store the 
bit vector $\langle{\tt 01101000}\rangle$, using 8 bits for 5 elements.

The index of an element of $Y$ is determined by querying the filter a
number of times, stopping when it returns $1$. In this example, to
determine the index of $y_3$, we first query the filter with $(y_3,
1)$ and obtain the hashes $4$ and $8$. 
Thus, the filter will return $0$ because it has {\tt 0} at the positions 4 and 8, 
and $0\oplus 0 \equiv 0$. Next, we query the filter with $(y_3,
2)$. The filter will again return $0$. Finally, we query the filter with
$(y_3, 3)$. Now it will return $1$. This means the MPHF index
of $y_3$ is $H_3(y_3) = 4$.





\subsection{Proof of Concept}

We provide some experimental results of building and
testing the $1.83n$ MPHF construction. To this end, we wrote a tool
that is composed of implementations of a minimum-weight bipartite
perfect
matcher\footnote{\url{https://github.com/jamespayor/weighted-bipartite-perfect-matching}}
and an XORSAT filter~\cite{Weaver18}.
All runs were performed on an Amazon EC2 instance
of type {\tt m5.4xlarge}. Such an instance has 16 cores, 64 GiB of memory,
and currently costs 0.768 USD per hour.

\begin{table}[ht]
\begin{center}
\caption{Runtime in seconds to compute the minimum-weight bipartite perfect
  matching on $n$ keys, per the construction described
  above.}\label{tab:mwbptimes} ~\\
\begin{tabular}{@{\,\,}c@{\,\,}c@{\,\,}c@{\,\,}c@{\,\,}c@{\,\,}c@{\,\,}c@{\,\,}c@{\,\,}c@{\,\,}c@{\,\,}c@{\,\,}}
\toprule
~$2^{9}$~ & ~$2^{10}$~ & ~$2^{11}$~ & ~$2^{12}$~ & ~$2^{13}$~ & ~$2^{14}$~ & ~$2^{15}$~ & ~$2^{16}$~ & ~$2^{17}$~ & ~$2^{18}$~ & ~$2^{19}$~ \\
\midrule
$<$$1$ & $<$$1$ & $<$$1$ & $<$$1$ & $<$$1$ & $1$ & $2$ & $9$ & $32$ & $134$ & $566$ \\
\bottomrule
\end{tabular}
\end{center}
\end{table}

The results in Table~\ref{tab:mwbptimes} show that it will not be
practical to build a large MPHF without blocking (or sharding) the
input and solving each block independently. The trade-off here is
that, in exchange for being able to build MPHFs for very large sets,
blocking slightly decreases the space efficiency of the MPHF due to
the need to store extra information relating to block
sizes. Table~\ref{tab:mwbptimes} also gives some insight into
appropriate block sizes for MPHFs: a block size between
$2^{13}$ and $2^{14}$ allows blocks to be built in about one second.

\begin{table}[htbp]
\centering
\caption{Achieved bits per key (BPK) and seconds taken to compute a blocked
  MPHF on $n$ keys, per the construction described above. Size refers to the number of kilobytes in the resulting
MPHF. Cost of building is in USD. Query speed is in nanoseconds per
query.}\label{tab:mphftimes} ~\\
{
\begin{tabular}{l@{~~~}r@{~~~}r@{~~~}c@{~~~}c@{~~~}c}
\toprule
&
\multicolumn{2}{c}{
   \begin{tabular}{c}
      Build Time
   \end{tabular}} & & & Query \\
$n$ & 1 Core & 16 Cores & BPK & Cost & Speed \\
\midrule
$2^{15}$ & $3$ & $<1$~ & 1.85 & 0.00 & 172 \\
$2^{16}$ & $6$ & $<1$~ & 1.85 & 0.00 & 168 \\
$2^{17}$ & $14$ & $2$~ & 1.85 & 0.00 & 165 \\
$2^{18}$ & $28$ & $3$~ & 1.85 & 0.00 & 167 \\
$2^{19}$ & $55$ & $4$~ & 1.85 & 0.00 & 167 \\
$2^{20}$ & $114$ & $8$~ & 1.85 & 0.00 & 167 \\
$2^{21}$ & $228$ & $15$~ & 1.85 & 0.00 & 167 \\
$2^{22}$ & $450$ & $30$~ & 1.85 & 0.01 & 170 \\
$2^{23}$ & $903$ & $59$~ & 1.85 & 0.01 & 191 \\
$2^{24}$ & $1\,816$ & $118$~ & 1.85 & 0.03 & 225 \\
$2^{25}$ & $3\,503$ & $228$~ & 1.85 & 0.05 & 211 \\
$2^{26}$ & $7\,238$ & $472$~ & 1.85 & 0.10 & 221 \\
$2^{27}$ & $14\,465$ & $952$~ & 1.85 & 0.20 & 244 \\
$2^{28}$ & $28\,991$ & $1\,920$~ & 1.85 & 0.41 & 335 \\
$2^{29}$ & $57\,861$ & $3\,892$~ & 1.85 & 0.82 & 420 \\
$2^{30}$ & $120\,088$ & $8\,346$~ & 1.85 & 1.78 & 552 \\
\bottomrule
\end{tabular}}
\end{table}


In Table~\ref{tab:mphftimes}, to maintain high XORSAT-filter
efficiency up into the billions of keys, the XORSAT-filter parameters
were raised from those suggested by Weaver et al.~\cite{Weaver18} to a
block size of $4608$. We also used the sparse vector generation method
proposed by Dietzfelbinger and Walzer~\cite{Dietzfelbinger19}. As
expected, each run generated a weighted bipartite graph with a minimum
weight of $1.831 n$, so any loss in space efficiency is
due to extra information stored about the MPHF blocking scheme and
inefficiencies in the resulting XORSAT filters.

Recent practical improvements to the construction of space-efficient
retrieval structures, such as those proposed by Dietzfelbinger and
Walzer~\cite{DBLP:conf/esa/DietzfelbingerW19}, would likely improve
the results given here (shorter build time) and we plan to evaluate
them in future work.

We performed a comparison between our approach and the
state-of-the-art. Specifically, we used the {\tt benchmphf}
tool\footnote{\url{https://github.com/rchikhi/benchmphf}} to generate MPHFs for $2^{30}$ elements. The {\tt
benchmphf} tool supports the {\tt
EMPHF}~\cite{DBLP:conf/dcc/BelazzouguiBOVV14}, {\tt
HEM}~\cite{Botelho13}, {\tt CHD}~\cite{Belazzougui09}, and {\tt
Sux4j}\footnote{\url{https://github.com/vigna/Sux4J}} implementations. Our results were similar to those
reported by Limasset et al.~\cite{Limasset17} with construction time
between $400$ and $2000$ seconds, bits per key between $2.90$ and
$4.22$, and query time between $250$ to $1000$ nanoseconds per
query. For $2^{30}$ elements, our implementation achieved $1.85$ bits
per key, took $120000$ seconds (sequentially) to build, and has a
query time of roughly $550$ nanoseconds per query.

\section{Perfect Hash Functions}

Our paper focuses on MPHFs, but the same techniques can
also be used to construct (non-minimal) perfect hash functions (PHFs). 
We already observed for the SAT-based approach that 
increasing the number of variables (i.e, enlarging the distance
with respect to the information-theoretic limit) reduces the costs to 
build a MPHF. Alternatively, one can reduce the costs by
increasing the range to which keys are mapped and thus allow
gaps. 

When considering the $1.83n$ MPHF construction reworked for PHFs, 
the ratio of right-hand-side nodes to left-hand-side nodes
in the bipartite graph will be greater than $1$. How does this effect
the minimum weight of the graph? If the weight scales slowly as the
ratio grows, the construction proposed here could also be used to build
space-efficient PHFs.

\section{Conclusions and Future Work}

We presented two MPHF constructions. Our first construction
achieves storage efficiency near the information-theoretic limit but has the
drawback that SAT-solving techniques can currently only deal with instances up to 40 keys,
which is still twice as much as previous (brute-force) approaches. 
Our second construction is practical in that it allows
constructing MPHFs for large datasets for low costs and high query
speeds. In fact, this is the most storage-efficient
construction of all currently known practical MPHF constructions.

One of the exciting challenges for future work is to find out whether a SAT-solving
tool can be developed that can construct MPHFs near the information-theoretic limit
for a substantial number of elements. Such an approach would likely be 
based on local-search techniques as they tend to scale better on
random problems, such as uniform random $k$-SAT formulas. 

Our experiments revealed the surprising observation that a huge encoding without auxiliary variables achieves better results than more compact encodings. As stated earlier, 
this may be due to the increased propagation power of the cubic encoding.
The development of an encoding with only $\mathcal{O}(n^2 \log_2 n)$ clauses
with the same propagation power as the cubic encoding could further improve
the results. 




\subsection{Acknowledgements}

The second author is supported by the National Science Foundation (NSF) under grant CCF-1813993.
The authors thank Benjamin Kiesl and the anonymous reviewers for their 
valuable input to improve the quality of the paper. 
The authors acknowledge the Texas Advanced Computing Center (TACC) at The University of Texas at Austin for providing HPC
resources that have contributed to the research results reported within this paper.


\bibliographystyle{aaai}
\bibliography{satbib}

\end{document}